\documentclass[12pt,twocolumn]{article}
\usepackage[letterpaper,margin=1in]{geometry}
\usepackage{pictex}
\usepackage[numbers]{natbib}
\usepackage{abstract}
\begin{document}
\title{Expectation of the Site Frequency Spectrum} 
\author{Alan R. Rogers\thanks{Dept.\ of Anthropology, 260 Central
    Campus Dr, University of Utah, Salt Lake City, UT
    84112. rogers@anthro.utah.edu} \and Stephen P.
  Wooding\thanks{Dept.\ of Anthropology, University of California Merced,
  5200 North Lake Road, Merced, CA 95343.}}
\twocolumn[
\maketitle
\begin{onecolabstract}
  The site frequency spectrum describes variation among a set of $n$
  DNA sequences. Its $i$'th entry ($i=1,2,\ldots,n-1$) is the number
  of nucleotide sites at which the mutant allele is present in $i$
  copies. Under selective neutrality, random mating, and constant
  population size, the expected value of the spectrum is well known
  but somewhat puzzling. Each additional sequence added to a sample
  adds an entry to the end of the expected spectrum but does not
  affect existing entries. This note reviews the reasons for this
  behavior.
\end{onecolabstract}
]
\saythanks

In a sample of $n$ DNA sequences, a polymorphic nucleotide site can
divide the sample into 1 mutant and $n-1$ non-mutants, into 2 mutants
and $n-2$ non-mutants, and so on. The number of copies of the mutant
must be at least 1 and no more than $n-1$ if the site is polymorphic.
The \emph{site frequency spectrum} describes the number of sites that
fall into each of these $n-1$ categories. It is widely used as a
summary of variation among DNA sequences.

The expectation of the spectrum has a very simple form under random
mating, selective neutrality, and constant population size, provided
that the mutation rate is so low that we can ignore the possibility of
multiple mutations at the same site---the so-called ``infinite sites''
model of mutation \citep{Kimura:InfiniteSites}. The expected spectrum
for samples of sizes 2 through 5 looks like this:
\begin{center}
  \begin{tabular}{cl}
Sample & Expected spectrum\\
size   & (singletons, doubletons, $\ldots$)\\ \hline
2 & $\theta$\\
3 & $\theta, \quad \theta/2$\\
4 & $\theta, \quad \theta/2, \quad \theta/3$\\
5 & $\theta, \quad \theta/2, \quad \theta/3, \quad \theta/4$\\
\multicolumn{2}{c}{Etcetera}
\end{tabular}
\end{center}
Here, ``singletons'' are sites with one copy of the mutant allele,
``doubletons'' are sites with two copies, and so on. $\theta=4Nu$,
where $N$ is population size and $u$ the mutation rate per sequence
per generation. The expected number of sites with $i$ copies of the
mutant allele is $\theta/i$ \citep[Eqn.~22]{Fu:TPB-48-172}. It is
remarkable that as we increase sample size, the number of mutants in
each category doesn't change.  We merely add a new category at the
end. To explore the cause of this behavior, we begin with a graphical
argument.

\begin{figure*}
  {\centering\input{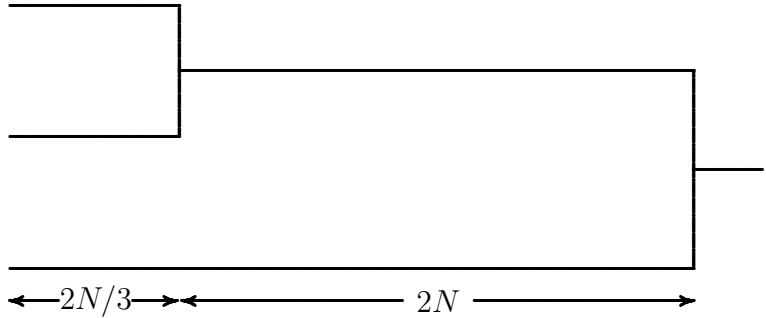}\\}
  \caption{Gene genealogy of a sample of size 3, with the expected
    length in generations of each coalescent interval.}
  \label{fig.tree3}
\end{figure*}

Figure~\ref{fig.tree3} shows the gene genealogy of a sample of size 3,
with branch lengths equal to their expected values
\citep{Hudson:OSE-7-1}. There are two coalescent intervals: a recent
one on the left, with three lines of descent, and an ancient one on
the right, with two. Imagine first that we could examine this sample
at the end of the ancient interval, just before the upper lineage
bifurcates. Our tree would have two branches, and the expected length
of each is $2N$ generations. Thus, we expect $2Nu = \theta/2$
mutations on each branch, and $\theta$ mutations altogether, all of
which are singletons.

At the time of the coalescent event, half of these singletons become
doubletons, and this is why we end up with $\theta/2$ doubletons at
the left edge of the graph. There are also new singletons, which arise
via mutation during the recent interval. How many? The expected length
of this interval is $2N/3$ generations, and there are 3 lines of
descent, so the total branch length within the interval is
$2N$. Consequently, the recent interval will add $\theta/2$ to the
expected number of singletons---exactly the number that we lost when
the upper lineage bifurcated. At the end of the second interval, we
expect $\theta$ singletons and $\theta/2$ doubletons, just as in the
table above.

To carry this argument farther, we must make it algebraic, and our
algebraic argument paraphrases that of
\citet{Hudson:PLO-10-e0118087}. As in the graphical argument, we
census mutations at the recent (tipward) end of each coalescent
interval.  For the $k$th interval (the one with $k$ lineages), let
$s_{i,k}$ denote the expected number of polymorphic sites at which the
derived allele is present in $i$ copies. There can be no polymorphic
sites if there is only one lineage, so $s_{i,1}=0$ for all $i$.
Singletons require special treatment.  Their expected number obeys
\begin{equation}
s_{1,k} = \frac{\theta}{k-1} + \left(1 - \frac{1}{k-1}\right)s_{1,k-1}
\label{eq.s1}
\end{equation}
The first term on the right accounts for singleton mutations that
arose within the current ($k$th) interval.  This is simply the number
of new mutations, a standard result from coalescent
theory.\footnote{To derive the expression, note that the expected
  duration of the interval is $4N/[k(k-1)]$.  Multiply by $k$ to get
  the total branch length and then by $u$ to get the expected number
  of mutations.}  The second term is the contribution from the
$s_{1,k-1}$ singletons that existed during the $(k-1)$th interval.  Of
these, a fraction $1-1/(k-1)$ remain singletons because they lie on a
lineage that did not bifurcate.

Consider now those mutations with $i>1$ copies at the end of interval
$k$. These can arise in two ways: either the mutation was present in
$i-1$ copies at the end of interval $k-1$, and one of the copies
bifurcated, or else it was present in $i$ copies, none of which
bifurcated. We consider these cases separately.

\emph{Mutations that increased in number.} During the interval with
$k-1$ lineages, $s_{i-1,k-1}$ is the expected number of mutations
present in $i-1$ copies. At the end of this interval, a single random
lineage bifurcates. Each lineage has the same chance, $1/(k-1)$, of
being the one that bifurcates, so a mutation with $i-1$ copies has
probability $(i-1)/(k-1)$ of including the lineage that bifurcates. In
this case, its number will increase from $i-1$ to $i$. Thus,
$\frac{i-1}{k-1}s_{i-1,k-1}$ is the contribution to $s_{i,k}$ from
mutations that increased in number.

\emph{Mutations that did not increase.} During interval $k-1$,
$s_{i,k-1}$ is the expected number of mutations present in $i$ copies.
With probability $1 - i/(k-1)$, none of these were on the lineage that
bifurcated. Thus, $(1 - i/(k-1))s_{i,k-1}$ is the contribution to
$s_{i,k}$ from mutations that did not increase.

Summing these contributions,
\begin{equation}
  s_{i,k} = \frac{i-1}{k-1}s_{i-1,k-1} + \left(1 -
  \frac{i}{k-1}\right)s_{i,k-1}
  \label{eq.si}
\end{equation}
for the case in which $i>1$.

Consider the sequence $s_{1,2}$, $s_{2,3}$, $s_{3,4}$, and so on. This
is the case of a ``lucky'' mutation whose count grows as fast as
possible, because one of its copies happens to bifurcate at the end of
each coalescent interval. It is of interest because it simplifies our
equations. With $k=i+1$, the second term in Eqns.~\ref{eq.s1}
and~\ref{eq.si} each disappear, leaving
\begin{eqnarray*}
s_{1,2} &=& \theta\\
s_{i,i+1} &=& \frac{i-1}{i}s_{i-1,i} \qquad \hbox{if $i>1$}
\end{eqnarray*}
If you work your way through this sequence, you will discover that
$s_{i, i+1} = \theta/i$ for all $i$.

Now we are in a position to discover why the entries of the expected
spectrum do not change as we add DNA sequences to the sample---or in
other words, why $s_{i,k}$ does not depend on $k$.  Rearrange
Eqns.~\ref{eq.s1} and~\ref{eq.si} as
\begin{eqnarray*}
  s_{1,k} &=& s_{1, k-1} + \frac{\theta - s_{1, k-1}}{k-1}\\
  s_{i,k} &=& s_{i,k-1} + \frac{(i-1)s_{i-1,k-1} - i s_{i,k-1}}{k-1}
\end{eqnarray*}
Using either equation, we can calculate $s_{i,i+2}$ from $s_{i,i+1}$,
then $s_{i,i+3}$ from $s_{i,i+2}$, and so on.  At each step the second
term disappears, leaving $s_{i,k} = s_{i, k-1}$.  This shows that
$s_{i,k}$ does not depend on $k$.  Its value is given by the formula
derived above for $k=i+1$:
\begin{equation}
s_{i,k} = \theta/i 
\label{eq.spec}
\end{equation}
for all $i$ and for all $k>i$, as shown by
\citet[Eqn.~22]{Fu:TPB-48-172}.

\bibliographystyle{plainnat}
\bibliography{defs,molrec,tree}

\end{document}